\documentclass{article}
\usepackage{amsfonts,graphicx,lscape}

\usepackage{amssymb}
\usepackage{eucal}
\usepackage{amsmath}
\usepackage{amsthm}
\usepackage{amscd}
\usepackage{graphics}
\usepackage{graphicx}
\usepackage{amsfonts}
\usepackage{latexsym}
\usepackage[all]{xy}

\newcommand{\R}{{\mathbb  R}}
\numberwithin{equation}{section}
\newtheorem{thm}{\bf Theorem}[section]

\newtheorem{proposition}[thm]{\bf Proposition}

\theoremstyle{remark}

\newcommand{\bc}{\begin{center}}
\newcommand{\ec}{\end{center}}
\newcommand{\bec}{\begin{equation}}
\newcommand{\eec}{\end{equation}}
\newcommand{\bea}{\begin{eqnarray}}
\newcommand{\eea}{\end{eqnarray}}
\newcommand{\ba}{\begin{array}}
\newcommand{\ea}{\end{array}}
\newcommand{\no}{\nonumber\\}

\newcommand{\f}{\displaystyle\frac}

\def\R{\mathbb{R}}
\def\ds{\displaystyle}
\def\vs{\vspace{4pt}}

\begin{document}

\title{On some dynamical and geometrical properties of the Maxwell-Bloch equations with a quadratic control}
\author{Tudor B\^inzar and Cristian L\u azureanu\\
{\small Department of Mathematics, "Politehnica" University of Timi\c soara}\\
{\small Pia\c ta Victoriei nr. 2, 300006 Timi\c soara, Rom\^ania}\\
{\small E-mail: tudor.binzar@upt.ro;cristian.lazureanu@upt.ro}}
\date{}

\maketitle

\begin{abstract}
\noindent In this paper, we analyze the stability of the real-valued
Maxwell-Bloch equations with a control that depends on state
variables quadratically. We also investigate the topological
properties of the energy-Casimir map, as well as the existence of
periodic orbits and explicitly construct the heteroclinic orbits.
\footnote{Journal of Geometry and Physics
Volume 70, August 2013, Pages 1–-8.\\http://dx.doi.org/10.1016/j.geomphys.2013.03.016}
\end{abstract}

\noindent \textbf{Keywords:} Maxwell-Bloch equations, Hamiltonian systems, stability
theory, energy-Casimir map, periodic orbits, heteroclinic orbits.

\section{Introduction}
The description of the interaction between laser light and a material sample composed of two-level atoms begins with
Maxwell$'$s equations of the electric field and Schr$\ddot{\mbox{o}}$dinger$'$s
equations for the probability amplitudes of the atomic levels. The resulting dynamics is given by Maxwell - Schr$\ddot{\mbox{o}}$dinger
equations which have Hamiltonian formulation and moreover there exists a homoclinic chaos \cite{HolKov92}.

Using the Melnikov method
\cite{Melnikov63}, in \cite{HoKoWe96} the presence of special homoclinic orbits for the dynamics of an ensemble of two-level atoms in a single-mode resonant laser
cavity with external pumping and a weak coherent probe modeled by Maxwell-Bloch's equations with a probe was established.

Fordy and Holm \cite{ForHol91} discussed the phase space geometry of the
solutions of the system introduced by Holm and Kovacic \cite{HolKov92}.

In 1992, David and Holm \cite{DavHol92} presented the phase space geometry
of the mentioned system restricted to $\R $, so named real-valued
Maxwell-Bloch equations:
\begin{equation}\label{eq1.1}
\left\{\begin{array}{l}
\dot x=y\\
\dot y=xz\\
\dot z=-xy
\end{array}\right.
\end{equation}

In 1996, Puta \cite{Puta96} considered system (\ref{eq1.1}) with a
linear, respectively a quadra-tic control $u$ about O$y$ axis:
\begin{equation}\label{eq1.2}
\left\{\begin{array}{l}
\dot x=y\\
\dot y=xz+u\\
\dot z=-xy
\end{array}\right.
\end{equation}
These particular perturbations arise naturally in controllability
context and were analyzed from the dynamical point of view. More
precisely, in the case of the quadratic control $u=(k-1)xz$ with
the parameter $k>0$ \cite{Puta96}, the dynamical analysis is done
by proving that the restricted dynamics on each symplectic leaf of
the associated Poisson configuration manifold is equivalent to the
dynamics of Duffing oscillator with control and with the pendulum
dynamics.

In our work, we consider system (\ref{eq1.2}), where $u=(k-1)xz$
with $k<0$. We give a Poisson structure and we find a symplectic
realization of the system. Using the method introduced in
\cite{TudAroNic09}, we find the image of the energy-Casimir map
and we study the topology of the fibers of the energy-Casimir map.

For details on Poisson geometry and Hamiltonian mechanical system,
see, e.g. \cite{CusBat97}, \cite{MarRat99}, \cite{Puta93},
\cite{LibMar87}.

\section{Poisson structure, symplectic realization and geometric prequantization}

Considering the quadratic parametric control $u=(k-1)xz$, system
(\ref{eq1.2}) becomes:
\begin{equation}\label{eq2.1}
\left\{\begin{array}{l}
\dot x=y\\
\dot y=kxz\\
\dot z=-xy
\end{array}\right.,
\end{equation}
where $k<0$ is the tuning parameter, according to the classification of chaos control methods \cite{CheDon93}, \cite{CheDon98}.

The constant of motion
$$H_{k}(x,y,z)=\f{1}{2}(y^2+kz^2)~,~~
C(x,y,z)=\f{1}{2}x^2+z$$
were given in \cite{Puta96}. Using the Euclidean space ${\R}^3$ with a modified cross-product as Lie algebra, a Poisson structure $\Pi $,
$$\Pi =\left[\begin{array}{ccc}0&1&\ds\ 0\\
-1&0&x\\-0&-x&0\end{array}\right]$$
was also given.

We are going to give a Lie algebra, isomorphic with that mentioned above, on its dual space the same Poisson structure is obtained.

Let us considering the Heisenberg Lie group $H_3$,
$$H_3=\{A\in GL(3,{\R})|~A=\left[\begin{array}{ccc}
1&a&c\\
0&1&b\\
0&0&1\end{array}\right],~a,b,c\in{\R}\}.$$

The corresponding Lie algebra $h_3$ is
$$h_3=\{X\in gl(3,{\R})|~X=\left[\begin{array}{ccc}
0&a&c\\
0&0&b\\
0&0&0\end{array}\right],~a,b,c\in{\R}\}.$$

Note that, as a real vector space, $h_3$ is generated by the
base\\
$B_{h_3}=\{E_1,E_2,E_3\},$ where
$$E_1=\left[\begin{array}{rrr}
0&0&1\\
0&0&0\\
0&0&0\end{array}\right],~E_2=\left[\begin{array}{rrr}
0&1&0\\0&0&0\\
0&0&0\end{array}\right],~E_3=\left[\begin{array}{rrr}
0&0&0\\0&0&1\\
0&0&0\end{array}\right].$$

The following bracket relations $[E_1,E_2]=0,[E_1,E_3]=0,[E_2,E_3]=E_1,$ hold.

Following \cite{LibMar87}, it is easy to see that the bilinear map
$\Theta :h_3\times h_3\to{\R}$ given by the matrix
$(\Theta_{ij})_{1\leq i,j\leq 3},$ $\Theta_{12}=-\Theta_{21}=1$
and 0 otherwise, is a 2-cocycle on $h_3$ and it is not a
coboundary since $\Theta (E_1,E_2)=1\not=0=f([E_1,E_2])$, for
every linear map $f,~f:h_3\to{\R}.$

On the dual space $h_3^*\simeq{\R}^3$, a modified Lie-Poisson structure is given in coordinates by
$$\Pi =\left[\begin{array}{ccc}0&0&0\\
0&0&x\\
0&-x&0\end{array}\right]+
\left[\begin{array}{rrr}
0&1&0\\-1&0&0\\
0&0&0\end{array}\right]=
\left[\begin{array}{ccc}0&1&0\\
-1&0&x\\
0&-x&0\end{array}\right].$$

The function $H_k$ is the Hamiltonian and $C$ is a Casimir of our configuration.

The next proposition states that system (\ref{eq2.1}) can be regarded as a Hamiltonian mechanical system.

\begin{proposition}
The Hamilton-Poisson mechanical system $(\R^3,\Pi ,H_k)$ has a
full symplectic realization $(\R^4,\omega,\tilde{H}_k)$, where
$$
\omega=dp_1\wedge dq_1 +dp_2\wedge dq_2
$$
and
$$
\tilde{H}_k=\ds\frac{1}{2}\left(p_1^2+kp_2^2-kp_2q_1^2+\frac{k}{4}q_1^4\right).
$$
\end{proposition}
{\bf Proof.}
The corresponding Hamilton's equations are:
\begin{equation}\label{eq2.2}
\left\{\begin{array}{l} \dot q_1=p_1 \vs\\
\dot q_2=kp_2-\ds\frac{k}{2}q_1^2 \vs\\ \dot p_1=kp_2q_1-\ds\frac{k}{2}q_1^3\vs\\
\dot p_2=0.
\end{array}\right.
\end{equation}
If we define the application $\varphi:\R^4\rightarrow\R^3$,
$$
\varphi(q_1,q_2,p_1,p_2)=(x,y,z)=\left(q_1,p_1,p_2-\ds\frac{1}{2}q_1^2\right),
$$
then it is easy to see that $\varphi$ is a surjective submersion, the equations (\ref{eq2.2}) are mapped onto the equations (\ref{eq2.1}) and
the canonical structure $\{\cdot,\cdot\}_\omega$ induced by $\omega$ is mapped onto the Poisson structure $\Pi $.

Therefore $(\R^4,\omega,\tilde{H}_k)$ is a full symplectic realization of the Hamilton-Poisson mechanical system $(\R^3,\Pi ,H_k)$.\\

Now, it is naturally to ask if system (\ref{eq2.2}) is completely integrable. The answer is given in

\begin{proposition}
The Hamilton mechanical system $(\R^4,\omega,\tilde{H}_k)$ given above is
completely integrable.
\end{proposition}
{\bf Proof.} Taking into account the definition of a completely
integrable Hamiltonian system \cite{LibMar87}, system (\ref{eq2.2})
has two differentiable first integrals
$$\tilde{H}_k=\ds\frac{1}{2}\left(p_1^2+kp_2^2-kp_2q_1^2+\frac{k}{4}q_1^4\right)~~\mbox{and}~~I=p_2$$
defined on $\R^4$, which are in involution
($\{\tilde{H}_k,I\}_\omega =0$), and whose differentials
are linearly independent on the dense open subset
\begin{eqnarray}
\Omega &=&\left\{(q_1,q_2,p_1,p_2)\in\R^4~:~\mbox{rank}\,J=2\right\}\nonumber\\
&=&\R^4\setminus\{(q_1,q_2,p_1,p_2)\,:\,p_1=0,\,q_1^3-2q_1p_2=0\}\nonumber
\end{eqnarray}
of $\R^4$, where $J$ is the Jacobian matrix of $\tilde{H}_k$ and $I$, which we set out to prove.\\

It is known that the symplectic manifold $(\R^4,\omega =d\theta )$, $\theta =p_1\,dq_1+p_2\,dq_2$, is quantizable from the geometric quantization
point of view \cite{Woodhouse} with the Hilbert representation space $\mathcal H=L^2(\R^4,\mathbb{C})$ and the prequantum operator
$\delta^\theta $,
$$\delta^\theta :f\in C^\infty(\R^4,\R)\mapsto\delta^\theta_f:\mathcal H\rightarrow\mathcal H,$$
where
$$
\delta^\theta_f=-i\hbar X_f-\theta (X_f)+f,
$$
$\hbar$ is the Planck constant divided by $2\pi$ and $X_f=\sum\limits_{k=1}^2\left(\ds\frac{\partial f}{\partial p_k}\cdot\ds\frac{\partial }{\partial q_k}-
\ds\frac{\partial f}{\partial q_k}\cdot\ds\frac{\partial }{\partial p_k}\right).$

We can state the following prequantization result:

\begin{proposition}
The pair $(\mathcal H,\delta)$, where $\mathcal H=L^2(\R^4,\mathbb{C})$ and
$$\delta:F\in C^\infty(\R^3,\R)\mapsto\delta_F~,~~\delta_F=\delta_{F\circ\varphi}^\theta ,$$
gives a prequantization of the Poisson manifold $(\R^3,\Pi )$.
\end{proposition}
{\bf Proof.} One easily check that Dirac's conditions are all satisfied.\\

We note that similar results for Maxwell-Bloch equations are given in \cite{Puta98}.

\section{Stability of equilibria and the image of the energy-Casimir mapping}
In this section we give the stability properties of the
equilibrium states of (\ref{eq2.1}) and we study the image of the energy-Casimir map $\mathcal E C_{k}$ associated with the Hamilton-Poisson
realization of system (\ref{eq2.1}).

The equilibrium states of system (\ref{eq2.1}) are given as the union of the following two families:
$$
\mathcal E_{k}^1=\left\{(M,0,0)\,|\,M\in\R\right\}
$$
$$
\mathcal E_{k}^2=\left\{\left(0,0,M\right)\,|\,M\in\R\right\}.
$$

We recall \cite{Puta96} that the equilibrium state $e_M=(M,0,0)\in \mathcal E_{k}^1,M\not=0,$ is unstable.

For the other equilibria we prove the following:
\begin{proposition}
Let $e_M=(0,0,M)\in \mathcal E_{k}^2$ be an arbitrary equilibrium state. Then $e_M$ is nonlinear stable for $M>0$
and unstable for $M<0.$
\end{proposition}
{\bf Proof.} For $M<0$, the eigenvalues of the characteristic polynomial associated with the linearization of system (\ref{eq2.1}) at
$e_M=\left(0,0,M\right)$ are $\lambda_1=0$, $\lambda_{2,3}=\pm \sqrt{kM}\in\R$, hence $e_M$ is unstable.

To study the nonlinear stability of the equilibria from $\mathcal E_{k}^2$ in the case $M>0$ we are using the Arnold stability test
\cite{Arnold65}, \cite{HoMaRaWei85}. To do that, let $F_\lambda\in{\mathcal{C}}(\R^3,\R)$ be defined by
$~F_\lambda =H_{k}-\lambda C=\f{1}{2}(y^2+kz^2)-\lambda\left(\f{1}{2}x^2+z\right) ,$ where $\lambda $ is a real parameter.
Then, we have successively the following:
\par (i) $~~\mbox{d}F_\lambda (0,0,M)=0$ if and only if $\lambda =kM$;
\par (ii) $~W=\ker\mbox{d}C(M,0,k)=\mbox{Sp}_{\R }\left\{\left(1,0,0\right),(0,1,0)\right\}$;
\par (iii) $\left.\mbox{d}^2F_{\lambda =kM}(0,0,M)\right|_{W\times W}=-kM\mbox{d}x^2+\mbox{d}y^2$ is positive definite for $M>0.$

Hence, from the Arnold stability test we conclude that all the equilibria from $\mathcal E_{k}^2$ with $M>0$ are nonlinear stable.\\

Now, let us consider the energy-Casimir mapping $\mathcal E C_{k}\in C^\infty(\R^3,\R^2)$, given by
$$
\mathcal E C_{k}(x,y,z)=\left(H_{k}(x,y,z),\,C(x,y,z)\right)=\left(\f{1}{2}(y^2+kz^2),\frac{1}{2}x^2+z\right).
$$

The next proposition gives a characterization of the image of the energy-Casimir map $\mathcal E C_{k}$.

\begin{proposition}
The image of the energy-Casimir map is $$Im(\mathcal E C_{k})=S_{k,-}^I\cup S_{k,-}^{II}\cup S_{k,-}^{III},$$
where \bea
S_{k,-}^I&=&\left\{(h,c)\in{\R}^2\,:\,c^2\leq
\f{2h}{k}\right\}\no S_{k,-}^{II}&=&\left\{(h,c)\in{\R}^2\,:\,c^2\geq \f{2h}{k},h\leq 0,c\geq 0\right\}\no
S_{k,-}^{III}&=&{\R}^2\setminus~\mbox{Int}~\left(S_{k,-}^I\cup
S_{k,-}^{II}\right). \nonumber \eea (Int$(A)=$ the interior of the
set $A$).
\end{proposition}
{\bf Proof.} The conclusion follows by algebraic computations using the definition of the energy-Casimir map.\\

As any semialgebraic manifold has a canonical Whitney stratification \cite{Pflaum01}, we will describe it by using the image through the
energy-Casimir map of subsets of the families of equilibria of system (\ref{eq2.1}).

\begin{proposition}
The semialgebraic canonical Whitney stratifications of $S_{k,-}^I$, $S_{k,-}^{II}$, $S_{k,-}^{III}$ are given by the following:

(i) $S_{k,-}^I=Im\left(\mathcal E C_{k}|_{\mathcal E_{k}^{2}}\right)\cup\Sigma^p(S_{k,-}^I)$,
where $\Sigma^p(S_{k,-}^I)=\{(h,c)\in\R^2:\,c^2<\frac{2h}{k}\}$ is the principal stratum.

(ii) $\ds{S_{k,-}^{II}}=Im\left(\mathcal E C_{k}|_{\mathcal E_{k}^{2,s}}\right)\cup
Im\left(\mathcal E C_{k}|_{\mathcal E_{k}^{1,u}\cap\mathcal E_{k}^{2,u}}\right)\cup Im\left(\mathcal E C_{k}|_{\mathcal E_{k}^{1,u}}\right)
{\cup\Sigma^p(S_{k,-}^{II})},$
where $\Sigma^p(S_{k,-}^{II})=\left\{(h,c)\in\R^2\,:\,c^2>\f{2h}{c},h<0,c>0\right\}$ is the principal stratum.

(iii) $\ds{S_{k,-}^{III}}\!=\! Im\left(\mathcal E C_{k}|_{\mathcal E_{k}^{2,u}}\right)\cup
Im\left(\mathcal E C_{k}|_{\mathcal E_{k}^{1,u}\cap\mathcal E_{k}^{2,u}}\right)\cup Im\left(\mathcal E C_{k}|_{\mathcal E_{k}^{1,u}}\right)
{\cup\Sigma^p(S_{k,-}^{III})},$
where the principal stratum $\Sigma^p(S_{k,-}^{III})$ is the interior of $S_{k,-}^{III}$.
\end{proposition}

All the stratification results can be gathered as shown in Figure 1, where we denoted $Im\left(\mathcal E C_{k}|_{{\mathcal E}_k^i}\right)$ by
$\Sigma_{k,-}^i$, and superscripts "$u$" and "$s$" stand respectively for "unstable" and "stable".
\begin{figure}[htbp]
\centering
\includegraphics[width=0.4\textwidth]{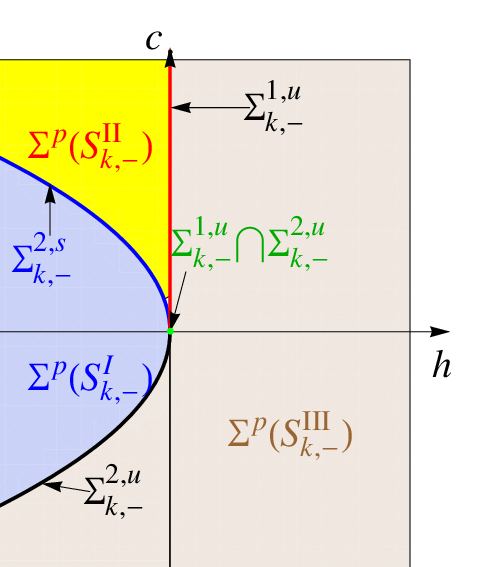}\label{Fig1}
\caption{Whitney canonical stratification of semialgebraic manifolds
$S_{k,-}^{I}$, $S_{k,-}^{II},~S_{k,-}^{III}$}
\end{figure}
\vspace*{10pt}

\section{The topology of the fibers of the energy-Casimir map}
In this section the topology of the fibers of $\mathcal E C_{k}$ is described.

A fiber of the map $\mathcal E C_{k}:\R^3\rightarrow\R^2$ is the preimage of an element $(h,c)\in\R^2$ through $\mathcal E C_{k}$, that is
$\mathcal F_{(h,c)}=\left\{(x,y,z)\in\R^3~:~\mathcal E C_{k}(x,y,z)=(h,c)\right\}.$

\begin{proposition}
The classification of the fibers $\mathcal F_{(h,c)}$ with $(h,c)$ belonging to the strata of $Im\left(\mathcal EC_{k}\right)$
can be described as follows:
\par (i) If $(h,c)\in\Sigma_{k,-}^{2,s}$ then $\mathcal F_{(h,c)}=\left\{(0,0,c):\,c>0\right\}\cup $\\
$\cup\left\{(x(t),y(t),z(t)):\,t\in (-\frac{\pi }{2\sqrt{-kM}},\frac{\pi }{2\sqrt{-kM}})\right\}\cup $\\
$\cup\left\{(-x(t),-y(t),z(t)):\,t\in (-\frac{\pi }{2\sqrt{-kM}},\frac{\pi }{2\sqrt{-kM}})\right\}$ (Figure 2), where
\bea
x(t)&=&2\sqrt{M}\sec (\sqrt{-kM}t)\no
y(t)&=&2M\sqrt{-k}\sec (\sqrt{-kM}t)\cdot\tan(\sqrt{-kM}t)\no
z(t)&=&-M[1+2\tan^2(\sqrt{-kM}t)].\nonumber
\eea
\par (ii) If $(h,c)\in\Sigma_{k,-}^{2,u}$ then\\
$\mathcal F_{(h,c)}=\{(x(t),y(t),z(t)):\,t\in (0,\infty )\}\cup\{(x(t),y(t),z(t)):\,t\in (-\infty,0)\}\cup $\\
$\cup\{(-x(t),-y(t),z(t)):\,t\in (0,\infty )\}\cup\{(-x(t),-y(t),z(t)):\,t\in (-\infty,0)\}$ (Figure 3), where
\bea
x(t)&=&\f{4\sqrt{Mp(t)}}{p(t)+1}\no
y(t)&=&\f{4M\sqrt{kp(t)}(p(t)-1))}{(p(t)+1)^2}\no
z(t)&=&M\cdot\f{p^2(t)-6p(t)+1}{(p(t)+1)^2}\nonumber
\eea
with $p(t)=-e^{2\sqrt{kM}t};$
\par (iii) If $(h,c)\in\Sigma_{k,-}^{1,u}$ then $\mathcal F_{(h,c)}$ is a pair of heteroclinic orbits (Figure 4);
\par (iv) If $(h,c)\in\Sigma^p(S_{k,-}^I)$ or $(h,c)\in\Sigma^p(S_{k,-}^{III})$ then\\
$\mathcal F_{(h,c)}=\{(x(t),y(t),z(t)):\,t\in (0,t_f)\}\cup\{(-x(t),y(t),z(t)):\,t\in (0,t_f)\}\cup $\\
$\cup\{(x(t),-y(t),z(t)):\,t\in (0,t_f)\}\cup\{(-x(t),-y(t),z(t)):\,t\in (0,t_f)\}$, where
\bea
x&=&\sqrt{2(c-z)}\no
y&=&\sqrt{2h-kz^2}\no
z&=&\f{2}{k}{\cal P}+\f{c}{3}\nonumber
\eea
where ${\cal P}$ is the Weierstrass elliptic function with the invariants $g_2=\f{k^2}{3}c^2+2kh$, $g_3=\f{k^3}{27}c^3-\f{2k^2}{3}hc$
(for details on elliptic functions see \cite{Lawden89});
\par (v) If $(h,c)\in\Sigma^p(S_{k,-}^{II})$ then $\mathcal F_{(h,c)}$ consists from the same four curves like as (iv) and a periodic orbit
(Figure 5).
\end{proposition}
{\bf Proof.} (i) Solving the system $(H_k(x,y,z),C(x,y,z))=(h,c)$, where $(h,c)\in\Sigma_{k,-}^{2,s}$ it results a solution $(0,0,c),~c>0$ and a
curve given by the intersection of surfaces $y^2+kz^2=2h$ and $x^2+2z=2c$ (Figure 2).

Considering the above surfaces as constants of motion we first reduce system (\ref{eq2.1}) from three degrees of freedom to one degree of freedom
and then integrate the resulting reduced differential equation, it follows the conclusion.

(ii)-(v) It results using the same method with $(h,c)$ from other stratum.\\

In the following we study the existence of periodic orbits of system (\ref{eq2.1}) around nonlinear stable equilibrium states.

Since the linearized system around nonlinear stable equilibrium states has a zero eigenvalue, we cannot apply Weinstein's result
\cite{Weinstein73} or Moser's result \cite{Moser76}
for proving the existence of periodic orbits. Then, we will apply Theorem 2.1 from \cite{BirPuTu07}, which ensures the existence of periodic orbits
around an equilibrium point. We recall this result:

"{\bf Theorem.} \emph{ Let $\dot{x}=X(x)$ be a dynamical system, $x_0$ an equilibrium point, i.e.,
$X(x_0)=0$ and $C:= (C_1,...,C_k): M\to\R^k$ a vector valued constant of motion for the
above dynamical system with $C(x_0)$ a regular value for $C$. If\\
(i) the eigenspace corresponding to the eigenvalue zero of the linearized system around
$x_0$ has dimension $k$,\\
(ii) $DX(x_0)$ has a pair of pure complex eigenvalues $\pm i\omega $ with $\omega\not =0$,\\
(iii) there exist a constant of motion $I:M\to\R$ for the vector field $X$ with $dI(x_0)=0$
and such that $d^2I(x_0)|_{W\times W}>0,$ where $W=\bigcap\limits_{i=1}^k\ker dC_i(x_0)$,\\
then for each sufficiently small $\varepsilon\in\R,$ any integral surface $I(x) = I(x_0) + \varepsilon^2$
contains at least one periodic solution of $X$ whose period is close to the period of the
corresponding linear system around $x_0$.}"\\

In our case, for $M>0$, the eigenvalues of the characteristic polynomial associated with the linearization of system (\ref{eq2.1}) at
$\left(0,0,M\right)$ are $\lambda_1=0$, $\lambda_{2,3}=\pm i\sqrt{-kM}$ and the eigenspace corresponding to the eigenvalue zero has dimension 1.
Considering the constant of motion $I:\R^3\rightarrow\R$,
$$
I(x,y,z)=-kMx^2+y^2+k(z-M)^2,
$$
it follows that $dI\left(0,0,M\right)=0$ and $\left.d^2I\left(0,0,M\right)\right|_{W\times W}>0$, where
$$W=\ker dC\left(0,0,M\right)=\mbox{Span}_{\R }\left\{(1,0,0),(0,1,0)\right\}.$$
By applying Theorem 2.1 from \cite{BirPuTu07} we have proved the following result:
\begin{center}
\begin{figure}[htbp]
\centering
\includegraphics[width=0.5\textwidth]{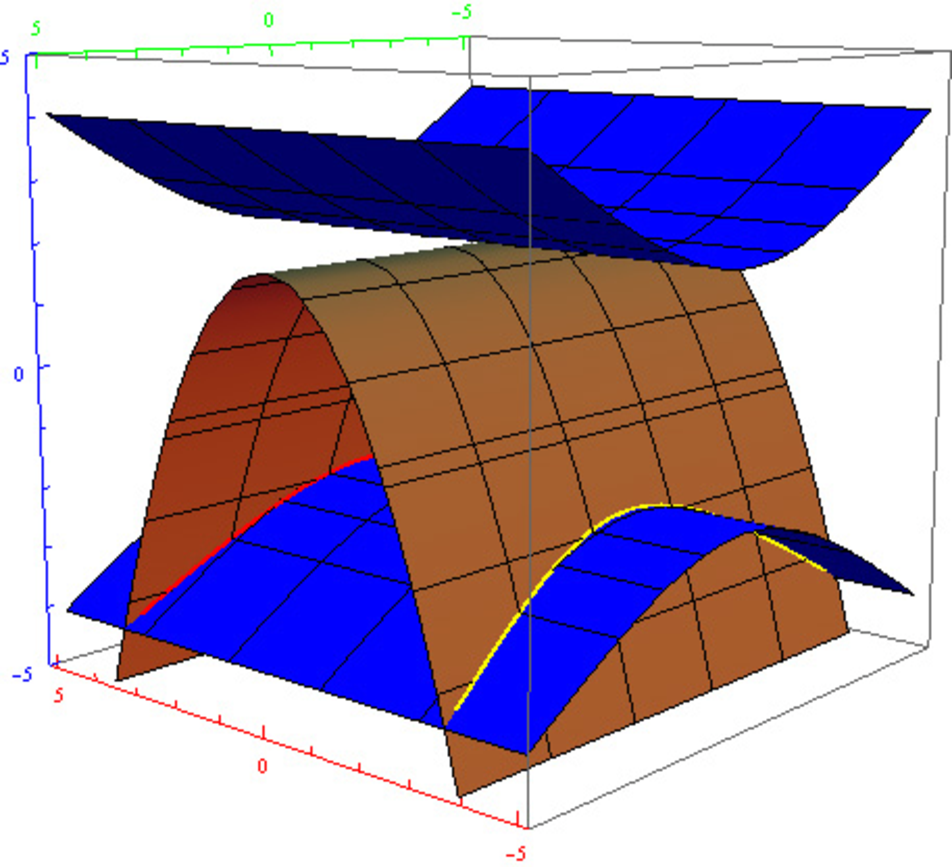}\label{Fig2}
\caption{The fiber $\mathcal F_{(h,c)}, (h,c)\in \Sigma_{k,-}^{2,s}$, consists of a stable equilibrium point and two curves, as intersection of two
surfaces}
\end{figure}
\end{center}
\begin{figure}[htbp]
\centering
\includegraphics[width=0.5\textwidth]{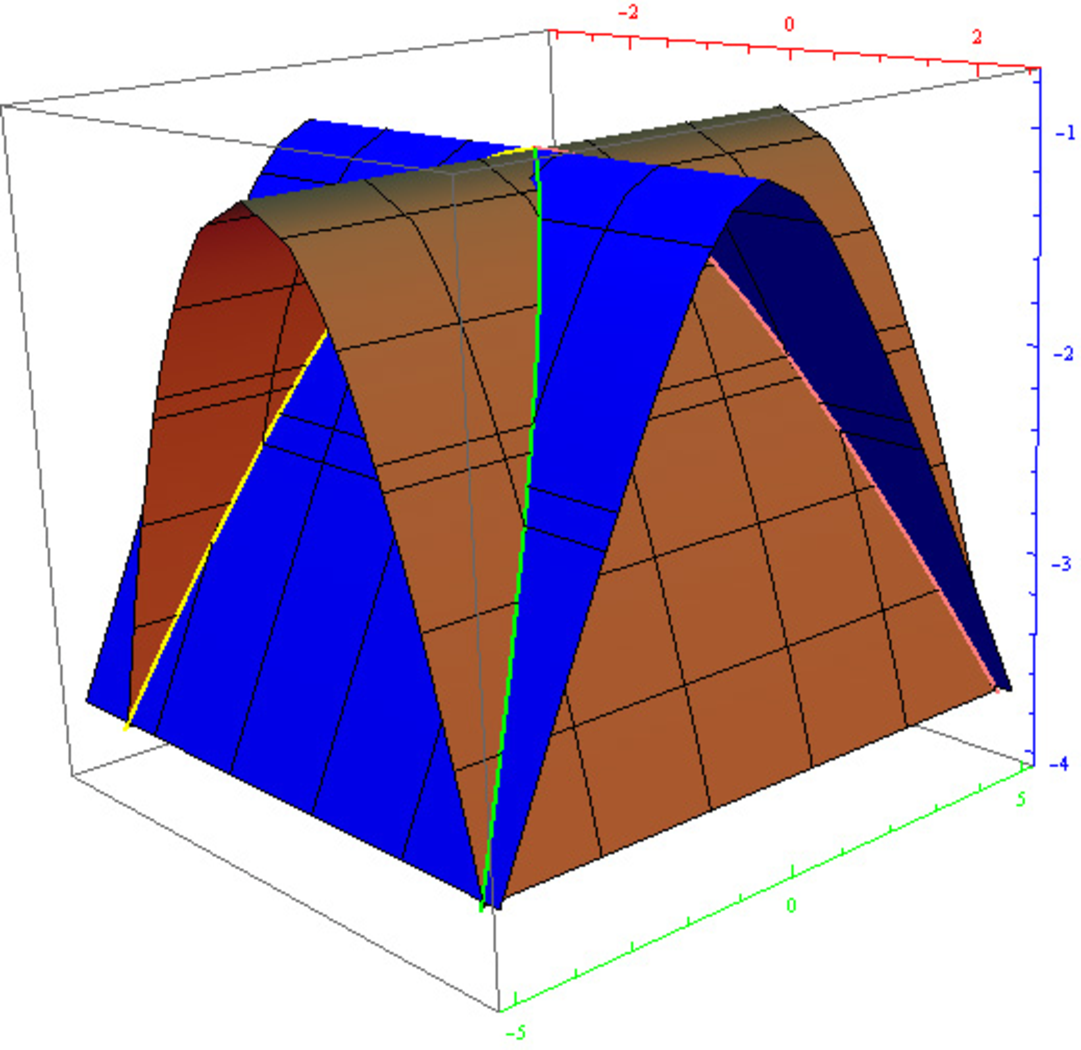}\label{Fig3}
\caption{The fiber $\mathcal F_{(h,c)}, (h,c)\in \Sigma_{k,-}^{2,u}$, consists of four curves, as intersection of two surfaces}
\end{figure}
\begin{figure}[htbp]
\centering
\includegraphics[width=0.5\textwidth]{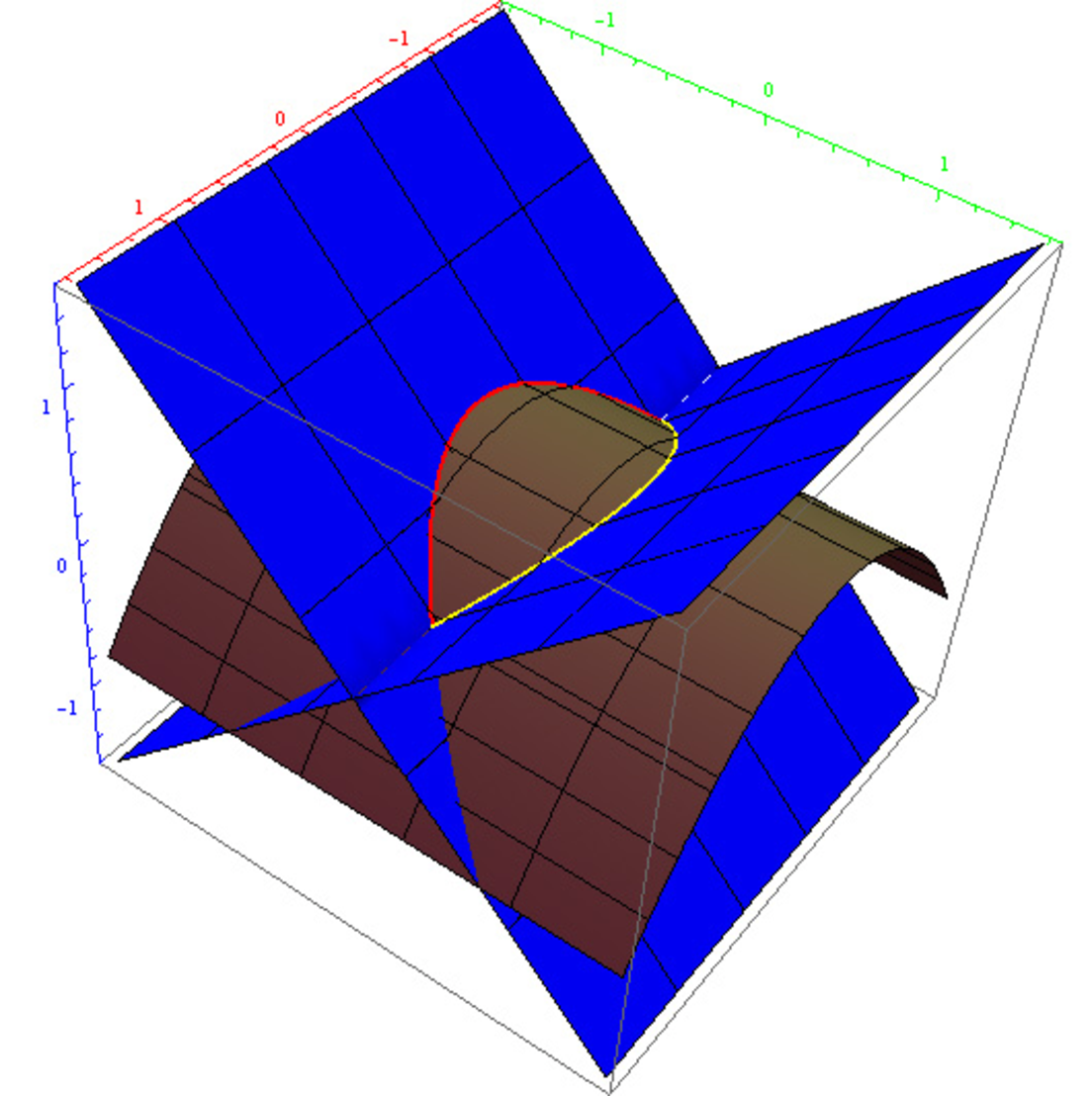}\label{Fig4}
\caption{The fiber $\mathcal F_{(h,c)}, (h,c)\in \Sigma_{k,-}^{1,u}$, consists of pair of heteroclinic orbits, as intersection of two surfaces}
\end{figure}
\begin{figure}[htbp]
\centering
\includegraphics[width=0.5\textwidth]{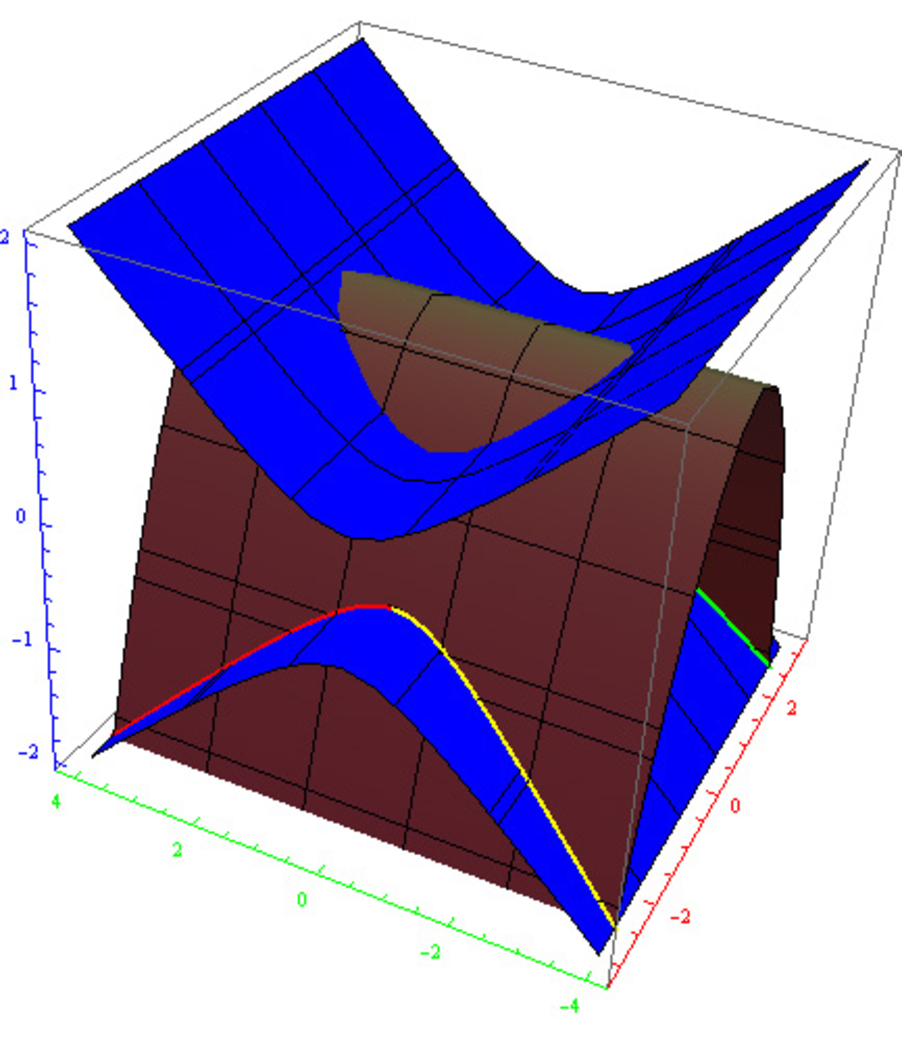}\label{Fig5}
\caption{The fiber $\mathcal F_{(h,c)}, (h,c)\in \Sigma^p(S_{k,-}^{II})$, consists of a periodic orbit and four curves, as intersection of two surfaces}
\end{figure}

\begin{proposition}
Let $e_M=\left(0,0,M\right)\in\mathcal E_{k}^{2,s}$ be such that $M>0$. Then for each sufficiently small
$\varepsilon\in\R_+^*$, any integral surface
$$
\Sigma_\varepsilon^{e_M}\,:\,~-kMx^2+y^2+k(z-M)^2=\varepsilon^2
$$
contains at least one periodic orbit $\gamma_\varepsilon^{e_M}$ of system (\ref{eq2.1}) whose period is close to
$\ds\frac{2\pi}{\sqrt{-kM}}$.
\end{proposition}\vs

In the sequel the heteroclinic orbits are given.

To obtain a parametric form of the heteroclinic orbits, we first reduce (\ref{eq2.1}) from three degrees of freedom to one degree of freedom
by using the relations obtained by eliminating $x$ and $y$ from $H_{k}(x,y,z)=H_{k}(0,0,M)$ and $C(x,y,z)=C(0,0,M)$
and then integrate the resulting reduced differential equation. We find $z$ and then $x$ and $y$. Thus, we obtain the following result:

\begin{proposition}
The parametrizations of the heteroclinic orbits connecting unstable equilibria\\ $(-M,0,0)$ and $(M,0,0)$, $M\not=0$, (Figure 4), are
\bea
\mathcal H_{(+,+,+)}^{(\pm M,0,0)}(t)&:=&(x(t),y(t),z(t)),\no
\mathcal H_{(-,-,+)}^{(\pm M,0,0)}(t)&:=&(-x(t),-y(t),z(t)),\nonumber
\eea
where
\bea
x(t)&=&M\cdot\f{p(t)-1}{p(t)+1}\no
y(t)&=&2M^2\sqrt{-k}\cdot\f{p(t)}{[p(t)+1]^2}\no
z(t)&=&2M^2\cdot\f{p(t)}{[p(t)+1]^2}\nonumber
\eea
with $p(t)=e^{M\sqrt{-k}(t+\alpha )},~~t\in\R $ and $\alpha\in\R$ is an arbitrary real constant.
\end{proposition}

\section*{Acknowledgments} We would like to thank the referees very much for their valuable comments and suggestions.


\begin{thebibliography}{99}

\bibitem{Arnold65} V. Arnold, Conditions for nonlinear stability of stationary plane curvilinear flows on an ideal fluid,
Akad. Nauk. Doklady SSSR, \textbf{162} (1965) 773--777.

\bibitem{BirPuTu07} P. Birtea, M. Puta, R.M.Tudoran,
Periodic orbits in the case of zero eigenvalue, C.R. Acad. Sci. Paris, Ser. I, \textbf{344} (2007) 779--784.

\bibitem{CheDon93} G. Chen, X.Dong, From chaos to order--Perspectives and methodologies in controlling chaotic
nonlinear dynamic systems, Int. J. of Bifurcation and Chaos, Vol. \textbf{3} (1993) 1363-–1409.

\bibitem{CheDon98} G. Chen, X.Dong, From Chaos to Order: Methodologies, Perspectives and Applications,
World Scientific Pub. Co., Singapore, 1998.

\bibitem{CusBat97} R.H. Cushman, L. Bates, Global Aspects of Classical Integrable Systems,
Birkh$\ddot{\mbox{a}}$user, Basel, 1997.

\bibitem{DavHol92} D. David, D.D. Holm, Multiple Lie-Poisson Structures, Reductions, and
Geometric Phases for the Maxwell-Bloch Travelling Wave Equations, J. Nonlinear Sci., Vol. \textbf{2} (1992) 241-262.

\bibitem{ForHol91} A. Fordy, D.D. Holm, A tri-Hamiltonian formulation of the self-induced transparency
equations, Phys. Lett. A, \textbf{160}(1991) 143-148.

\bibitem{HolKov92} D.D. Holm, G. Kovacic, Homoclinic chaos in a laser matter system, Physica D.,
Vol. \textbf{56} (1992) 270--300.

\bibitem{HoKoWe96} D.D. Holm, G. Kovacic, T.A. Wettergren, Homoclinic orbits in the Maxwell-Bloch
equations with a probe, Phys. Rev. E, \textbf{54} (1996) 243--256.

\bibitem{HoMaRaWei85} D.D. Holm, J. Marsden, T.S. Ra\c tiu, A. Weinstein, Nonlinear stability of
fluid and plasma equilibria, Physics Reports, vol. \textbf{123}, no. 2 (1985) 1--116.

\bibitem{Lawden89} D.F. Lawden, Elliptic Functions and Applications,
Springer--Verlag, New York, Berlin, Heidelberg, London, Paris, Tokyo, Hong Kong, 1989.

\bibitem{LibMar87} P. Libermann, C.-M. Marle, Symplectic Geometry and Analytical Mechanics, D. Reidel, Dordrecht, 1987.

\bibitem{MarRat99} J. Marsden, T.S. Ra\c tiu, Introduction to Mechanics and Symmetry, 2nd Ed.
Text and Appl. Math. \textbf{17}, Springer, Berlin, 1999.

\bibitem{Melnikov63} V.K.Melnikov, On the stability of the center for time periodic perturbations, Trans.
Moscow Math., \textbf{12} (1963) 1--57.

\bibitem{Moser76} J. Moser, Periodic orbits and a theorem by Alan Weinstein, Comm. Pure Appl. Math., \textbf{29} (1976) 727--747.

\bibitem{Pflaum01} M.J. Pflaum, Analytic and Geometric Study of Stratified Spaces, Lecture Notes in Math., \textbf{510},
Springer, Berlin, 2001.

\bibitem{Puta93} M. Puta, Hamiltonian mechanical system and geometric quantization,
Kluwer Academic Publishers, Dordrecht, Boston, London, 1993.

\bibitem{Puta96} M. Puta, Three-dimensional real-valued Maxwell-Bloch equations with controls,
Reports on Mathematical Physics, Volume \textbf{37}, Issue \textbf{3} (1996) 337--348.

\bibitem{Puta98} M. Puta, Integrability and geometric prequantization of the Maxwell-Bloch equations,
Bull. Sci. math, Volume \textbf{122} (1998) 243--250.

\bibitem{TudAroNic09} R.M. Tudoran, A. Aron, \c S. Nicoar\u a, On a Hamiltonian Version of the Rikitake System,
SIAM J. Applied Dynamical Systems, Vol. \textbf{8}, No. \textbf{1} (2009) 454--479.

\bibitem{Weinstein73} A. Weinstein, Normal modes for non--linear Hamiltonian systems, Invent. Math., \textbf{20} (1973) 47--57.

\bibitem{Woodhouse} N.M.J. Woodhouse, Geometric quantization, Oxford University Press, 1990.
\end{thebibliography}
\end{document}